\documentclass[a4paper,11pt]{article}
\usepackage{wrapfig}
\usepackage[format=plain,
            labelfont=it,
            textfont=it]{caption}
            
\usepackage{pos}
\usepackage{lipsum} 

\title{Angular dependence of the atmospheric neutrino flux with IceCube data}

\ShortTitle{Angular dependence of the atmospheric neutrino flux}

\author{The IceCube Collaboration \\{\normalsize \normalfont(a complete list of authors can be found at the end of the proceedings)}\\}

\emailAdd{leonora.kardum@tu-dortmund.de}

\abstract{

IceCube Neutrino Observatory, the cubic kilometer detector embedded in ice of the geographic South Pole, is capable of detecting particles from several GeV up to PeV energies enabling precise neutrino spectrum measurement. The diffuse neutrino flux can be subdivided into three components: astrophysical, from extraterrestrial sources; conventional, from pion and kaon decays in atmospheric Cosmic Ray cascades; and the yet undetected prompt component from the decay of charmed hadrons. A particular focus of this work is to test the predicted angular dependence of the atmospheric neutrino flux using an unfolding method. Unfolding is a set of methods aimed at determining a value from related quantities in a model-independent way, eliminating the influence of several assumptions made in the process. In this work, we unfold the muon neutrino energy spectrum and employ a novel technique for rebinning the observable space to ensure sufficient event numbers within the low statistic region at the highest energies. We present the unfolded energy and zenith angle spectrum reconstructed from IceCube data and compare the result with model expectations and previous measurements.

\vspace{4mm}
{\bfseries Corresponding authors:}
Leonora Kardum$^{*}$\\
{$^{}$ \itshape Department of Physics, TU Dortmund University, D-44221 Dortmund, Germany}\\
$^*$ Presenter

\ConferenceLogo{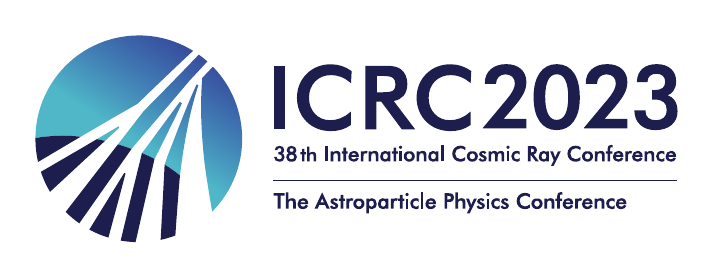}

\FullConference{The 38th International Cosmic Ray Conference (ICRC2023)\\ 26 July -- 3 August, 2023\\ Nagoya, Japan}
}

\begin{document}

\maketitle

\section{Introduction}\label{sec1}
The energy spectrum of neutrinos is a fundamental aspect of their properties and holds valuable information about their sources and interactions. Measurements of neutrinos generated in Cosmic Ray interactions with the atmosphere and astrophysical neutrinos therefore presents an unparalleled opportunity for neutrino research. The energy spectrum, spanning from GeV to PeV, comprises contributions from the conventional and prompt atmospheric neutrino components, as well as the astrophysical component at higher energies. The astrophysical component is measured to exhibit a harder spectrum, originating from astrophysical sources \cite{42}. However, this component has not been fully characterized, and as a result this particular energy range holds special significance and captures considerable interest.
\par
This work presents the unfolding method and its application in the measurement of the diffuse muon neutrino flux obtained from data taken with the IceCube Neutrino Observatory. The measurements are considered both regarding their energy dependence and zenith dependence.

\section{The IceCube Neutrino Observatory and its Data}
The IceCube Neutrino Observatory is a state-of-the-art detector located at the geographic South Pole \cite{icecube}. The observatory consists of an array of optical sensors called Digital Optical Modules (DOMs), which are embedded deep within the Antarctic ice spanning from 1.5 kilometers to 2.5 kilometers underground. Each of the 5160 DOMs houses a Photomultiplier Tube (PMT) and are subjected on 86 strings with regular spacing of 125 meters. DOMs capture the Cherenkov radiation produced by charged particles exceeding the speed of light in ice which is subsequently used to reconstruct 3-dimensional events of these secondary particles.  
\par
To reconstruct the flux of muon neutrinos, a subset of IceCube events referred to as track-like is necessary. Muons create long and straight signatures compared to electrons which are subject to higher energy loss during their propagation in ice. In this work, a subset of track-like events measured in IceCube over eleven years of data taking is aimed to be unfolded, with 850000 expected events between 500 GeV and 6 PeV. Utilizing Boosted Decision Trees (BDTs) to separate background, the sample is cleaned twice considering both the rejection of cascade events coming from electron neutrinos and the rejection of atmospheric muons not resulting from neutrino interactions \cite{40}. 
The meticulous process results in a notable purity of 99.7\%, with the energy resolution of $ log_{10} \frac{E_{rec}}{E_\mu} \approx 0.3$ at 100 TeV where angular resolution is 0.25 degrees. A preliminary analysis dataset referred to as $Burnsample$ is comprised of 10\% of the events, randomly selected from the total lifetime to be considered. 
\par
To train the response of the algorithm, a Monte Carlo simulation of neutrinos propagated through ice is used, with the simulation covering the energy range from 100 GeV to 500 PeV.   

\section{Muon neutrino flux and angular dependence}
The aforementioned conventional component of the neutrino flux is generated through Cosmic Ray interactions with nuclei in the Earth's atmosphere, producing charged particles, such as pions and kaons, which subsequently decay into neutrinos. The prompt neutrinos, on the other hand, are also produced in Cosmic Ray interactions but originate from the decay of short-lived particles usually containing the charm quark. The much shorter lifetime of charmed mesons reduces the interaction probability and these particles almost always decay before interacting. As a result of lower energy loss in the atmosphere, they contribute with a harder energy spectrum $\frac{d\Phi_p}{dE} \propto E^{-2.6}$, mirroring the spectral behaviour of the Cosmic Ray flux. The prompt component becomes substantial at energies around 100 TeV, and 
consequently, both conventional and prompt neutrinos are observed together and are collectively referred to as atmospheric neutrinos. The prompt neutrino component is yet to be observed and characterized \cite{prompt}, with recent results implicating the existence of prompt leptons \cite{promptleptons}. 
\par
Upon muon creation, which retains the direction due to the high energies of boosted primaries, the column depth of the atmosphere will impact its decay probability. When the angle of incidence is increased, the distance between the origin of the muons in the atmosphere and the detector increases, reaching its maximum at 180$^\circ$. For particles entering the atmosphere vertically, the interaction probability increases due to the higher density of particles at lower heights of the atmosphere. In contrast, entering at high angles produces a path with longer segments in lower air densities. Here, a lower interaction rate allows lower energy loss and consequently creates neutrinos of higher energies. Therefore, the conventional neutrinos from light mesons exhibit zenith-dependence in the energy spectrum. Contrary to the conventional component, the prompt particles do not spend sufficient time in the atmosphere to be influenced by these effects.
\par
The highest energy component is constituted of neutrinos from astrophysical sources, independent of processes on Earth and inside its atmosphere. Recent measurements indicate a spectral index $\frac{d\Phi_a}{dE} \propto E^{-2.62}$ for the astrophysical component \cite{42}, determined experimentally without a universally agreed-upon theoretical model to date. Due to the absence of a limiting horizon, the astrophysical component is expected to be isotropic which is consistent with current observations \cite{isotropy}. 

\section{Unfolding}
Unfolding refers to a set of techniques based on the principle of deconvolution, which aims to recover the underlying physical distribution of a source from observed data often influenced by instrumental and statistical effects. In the field of astrophysics, unfolding plays a crucial role in reconstructing the energy spectrum of a source, giving the radiation intensity as a function of particle energy. In this work, the focus lies on studying the diffuse neutrino flux, which characterizes the intensity of neutrinos reaching Earth from all directions and from unspecified sources. Unfolding involves the modeling of both the source and instrument properties, alongside precise estimation of the statistical and systematic uncertainties associated with the collected data.
\par
The event spectrum $f(x)$ is a function of neutrino energy $x$ and is distorted during the stochastic processes involved in neutrino detection. The smearing of the true distribution to the distribution measured in IceCube is described by a migration matrix $A$ which oughts to map the complete detection process, from the propagation and interactions of primary neutrinos and emerged leptons to the detector response described by measured features. The reconstructed energy is additionally hindered by the unknown amount of energy that the muon has lost before entering the detector volume. The measured distributions $g(y)$ after distortion are given by 
\begin{equation} \label{eq:unfolding}
\vec{g} =A_{m,n} \cdot \vec{f}
\end{equation}
where $g$ is often referred to as observable space. The sought-after target space $f(x)$ cannot simply be inferred by inverting the expression \ref{eq:unfolding} as the inversion of $A$ leads to unstable results. This is the direct result of ill-conditioning, the inversion of matrices with high condition numbers producing large changes in output caused by a light change in input data. 
\par
To mitigate these effects, the event spectrum per each considered energy range in the discretized space $f$ is reconstructed by maximising the defined Poissonian likelihood 
\begin{equation}
    \alpha (\vec{g} | \vec{f}) = \prod^m_{u=1} \frac{\lambda_u ^{g_u}}{g_u!} \cdot \exp(-\lambda_u)
\end{equation}
where the expected value for $g$ is given with Eq. (\ref{eq:unfolding}). Deriving the logarithm and simplifying leads to the expression for likelihood per each energy bin to be 

\begin{equation} \label{eq:likelihood}
    \alpha (\vec{g} | \vec{f}) = \sum ^m_{u=1} (g_u \ln (A\vec{f})_u - (A\vec{f})_u) ) - \frac{1}{2} \left [ (C\vec{f})^T Diag( \mathbb{I} \cdot \tau)  (C\vec{f}) \right ]
\end{equation}

where the subtrahend is the added regularization. Matrix $C$ corresponds to Tikhonov 
\begin{wrapfigure}{r}{0.5\textwidth}
\setlength{\intextsep}{0pt}%

    \begin{center}    
        \includegraphics[width=0.5\textwidth]{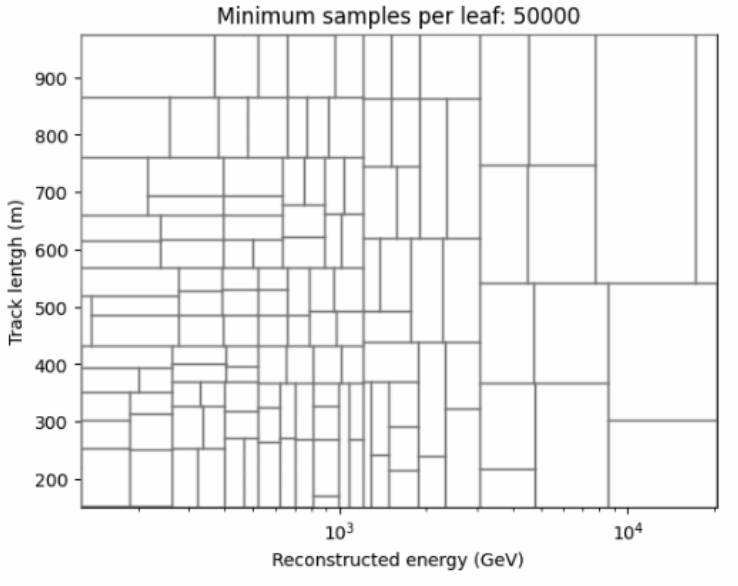}
        \caption{Observable space bins of the reconstructed track length (y-axis) depending on reconstructed energy (x-axis). Each of the example bins is populated with at least 50000 entries, ensured with the minimum leaf size. Higher bin counts correspond to regions with higher information gain. Here, the track length is densely branched in the low-energy range where the dominant energy loss mechanism is ionization. At low energies, the starting energy is correlated to track length, and lose importance in energy reconstruction as the dominant process turns to pair production in higher energy region.}
        \label{fig:rebinning}
        \vspace{-70pt}
    \end{center}
\end{wrapfigure}
regularization, a specific form commonly used in inverse problems, where the goal is to reconstruct an unknown function from noisy or incomplete data. This ensures smooth solutions by incorporating assumptions about the sought result. Due to the nature of the neutrino flux, the solution is expected to be a combination of at least three power-law spectra with different slopes. The additional term introduces a prior of the second derivative being close to zero, a property of power-law functions, and the stability is adjusted through the regularization strength parameter $\tau$. Regularized unfolding has been introduced with the algorithm RUN \cite{RUN}, has been used with the improved version named TRUEE \cite{truee} to achieve the unfolded muon neutrino energy spectrum \cite{ic79} and in a variation similar to one presented here \cite{mathis}.

\section{Rebinning the observable space}
Discretizing the observable space $g$ from Eq. (\ref{eq:unfolding}) into equidistant bins can distort the analysis for several reasons. Equally dividing data into bins without considering the distribution's imbalance can cause significant variations in statistics across those bins. When more features are added, the number of bins increases exponentially, resulting in reduced statistical accuracy, particularly in regions known to exhibit this property, like the high-energy region of the neutrino flux. Bins exceeding the resolution of the detector by several orders of magnitude lead to loss of information. Therefore, we present an approach to data preparation considering the distribution of each used observable to mitigate the stated effects. 
\par
A Decision Tree is trained to classify events into discretized energy bins. However, the classification done by the Tree is not used for energy reconstruction. While building the decision space, Trees optimize cuts with the goal of ensuring maximum information gain at each node. The leafs, final cuts in some set of decisions, are used as the bin edges in the new observable space. Setting the maximum number of leafs at 2500 sets an upper limit to the size of the observable space, and mandating a minimum of events needed to build a leaf at 100 ensures substantial statistic is available in each bin. Binning scheme is shown in Figure \ref{fig:rebinning}.

\section{The unfolded muon neutrino flux}
\begin{wrapfigure}{L}{0.5\textwidth}
\setlength{\intextsep}{0pt}%
    \begin{center}    
        \includegraphics[width=0.5\textwidth]{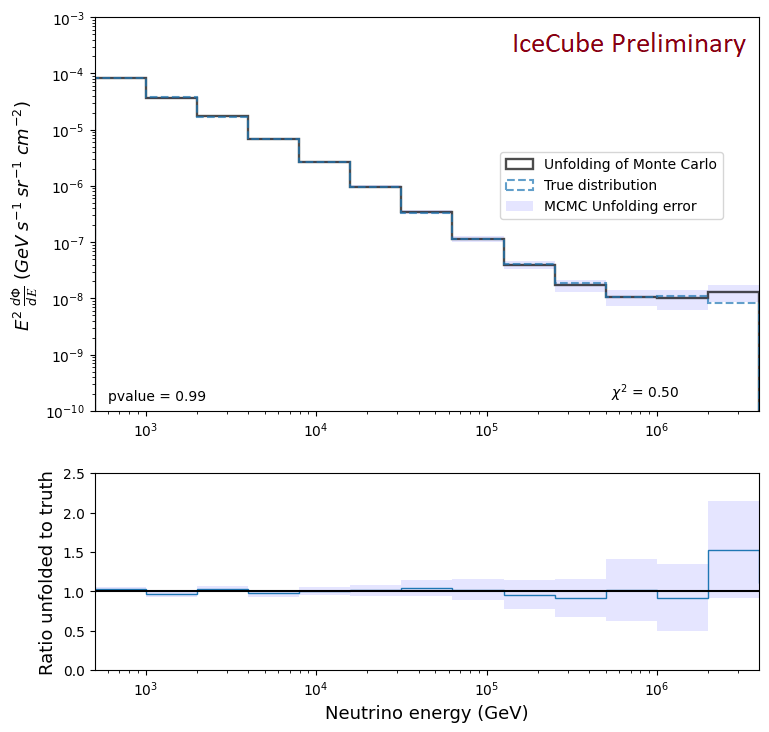}
        \caption{Unfolded weighted flux of a pseudosample shown in solid black line. The true distribution is shown in dashed blue lines and is always encompassed by the statistical errors shown in shaded area around the best estimate. The lower plot shows the ratio of the unfolded flux to the true value. Agreement between unfolded and true spectra are tested with the $\chi^2$ test, and the values are shown on the bottom left and right of the upper plot. }
        \label{fig:pseudounfolding}
        \vspace{-8pt}
    \end{center}
\end{wrapfigure}

To test the consistency of the procedure, pseudosamples, random sets imitating some underlying distribution (in this case the muon neutrino flux), are created and used to optimize and verify steps. In this work, pseudosamples are sampled assuming a primary Cosmic Ray flux model H4a \cite{Gaisser} and the interaction model Sybill2.3c \cite{Sybill} for the atmospheric components, and the astrophysical component taken from experimental results \cite{41}. For the given lifetime, this model predicts 655660.77 atmospheric events and 1147.25  of astrophysical origin. 
\par
Using the neutrino simulations as the training set, the migration matrix $A$ is populated into the observable space $\Vec{g}$ of size 2500. Markov Chain Monte Carlo (MCMC) walkers are employed to sample viable solutions of $\Vec{f}$ by calculating the likelihood given in Eq. (\ref{eq:likelihood}). Any movements that improves the likelihood is added to the distribution of estimates done by 10000 walkers for each energy bin. The energy range from 500 GeV to 6 PeV is divided into 13 bins in logarithmic space. The unfolded spectrum is compared to its true value by the means of $\chi^2$ test. 
\par
Unfoldings of all simulated pseudosamples show good agreement of the unfolded spectrum with its true value, with high significance as indicated through $\chi^2$ p-values indicated with each result. 
\par
Using the same optimization parameters and 9 features chosen based on their relevance and agreement to data, the algorithm is applied to a subset of data measured with IceCube, corresponding to 1 year, 20 days and 6 hours of lifetime. Unfolding of data is compared to current theoretical models for the atmospheric component, and several experimental results parametrizing the astrophysically dominated region. Atmospheric models are simulated with MCEq \cite{mceq}. The unfolding is in agreement with recent measurements, with bigger uncertainty due to the limited statistics in the $Burnsample$.

\begin{figure}[h!]
    \begin{center}    
        \includegraphics[width=0.8\textwidth]{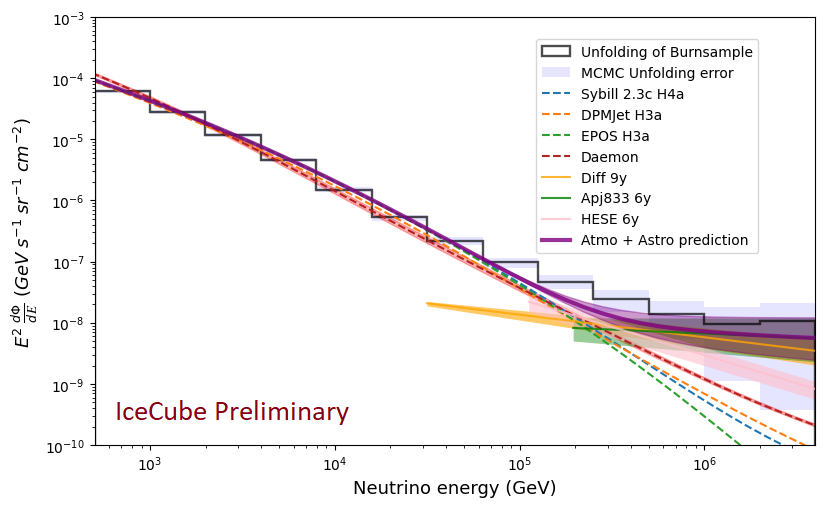}
        \caption{Weighted flux of the $Burnsample$ data. Black solid line shows the unfolding in this work. The dashed lines show different atmospheric models, with varying interaction models including Sybill2.3c \cite{Sybill} (blue), DPMJet \cite{DPMJet} (orange), and EPOS \cite{EPOS} (green), all with the Hillas-Gaisser primary Cosmic Ray flux model \cite{Gaisser}, and Daemonflux \cite{daemon}. Recent results from IceCube for the astrophysical components are shown in solid lines in their corresponding energy ranges, with the shading showing uncertainty. Included are the 6 years astrophysical fit \cite{41} (green), HESE analysis \cite{43} (pink), and the 9.5 years diffuse analysis \cite{42} (orange). The solid purple line shows the best estimate of the three-component neutrino flux by the combination of the astrophysical fit in green, and the atmospheric model in blue. Shaded blue area shows the uncertainty of unfolding around the best estimate.}
        \label{fig:burnsampleunfolding}
    \end{center}
\end{figure}
\par
Good agreement can be noticed in Figure \ref{fig:burnsampleunfolding} except for the unfolding between energies of 100 TeV and 400 TeV. A possible source of the disagreement is the aforementioned absence of precise prompt parametrization. Slight changes in the assumption of the power-law describing prompt neutrinos create substantial differences in the shape of the flux at this energy range, due to the crossover energy of prompt and conventional components being around 300 TeV \cite{gaisserprompt}, but not precisely known.
\par
The data is split into five angular subsets based on the expected event rates. The horizon is substantially more populated in events in comparison to the highest angles, in which the neutrinos have passed the whole diameter of Earth before reacting in the ice volume. From the simulation of the assumed atmospheric and astrophysical model, the event spectrum is divided into five angular bins of approximately the same number of events to ensure substantial statistics. The considered angular ranges are $86^\circ$ to $95^\circ $, 
$95^\circ$ to $105^\circ $, 
$105^\circ$ to $117^\circ $, 
$117^\circ$ to $134^\circ $, and
$134^\circ$ to $180^\circ $.
\par
\begin{wrapfigure}{L}{0.5\textwidth}
\setlength{\intextsep}{0pt}%
    \begin{center}    
        \includegraphics[width=0.5\textwidth]{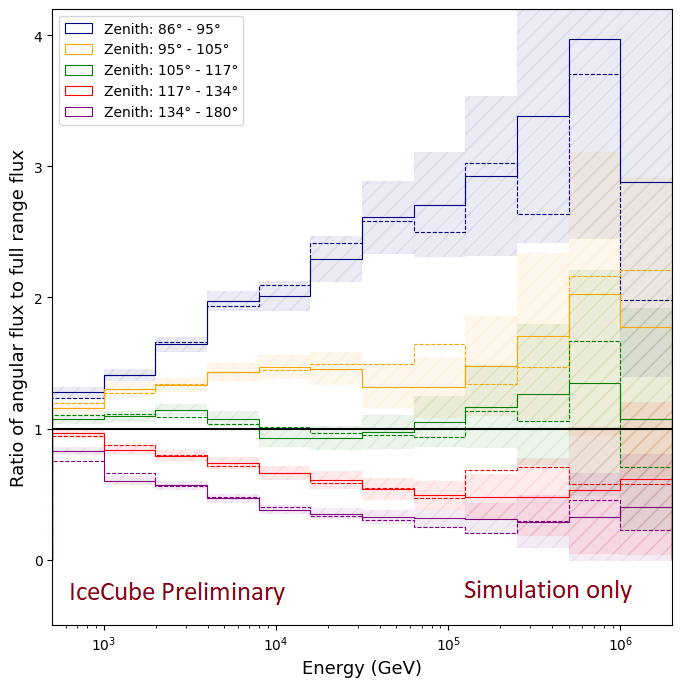}
        \caption{Angular unfolding of a pseudosample. The solid lines show the ratio of the unfolded flux in angular bins to the all sky unfolding. Dotted lines show the true values of the pseudosets sampled from the combined atmospheric prediction \cite{Sybill} and astrophysical fit \cite{41}. The horizon bin (blue) exhibits highest anisotropy stemming from the dominance of conventional particles. }
        \label{fig:pseudoangular}
        \vspace{-0pt}
    \end{center}
\end{wrapfigure}

Angular unfolding exhibits slight change in the size of uncertainty, due to a lower number of events in angular intervals compared to the full range. The angular fluxes can be discriminated up to energies of 1 PeV, where the error regions start to overlap, as shown in Figure \ref{fig:pseudoangular}.

\section{Discussion and Outlook}
We have presented the current state of unfolding the neutrino flux from 10\% of a dataset spanning eleven years of IceCube measurements. The algorithm shows promising results with low statistical error even in the low-statistic high energy region, as seen in the errors shown on figure \ref{fig:pseudounfolding}. The $Burnsample$ unfolding shows good agreement with theoretical predictions in the conventional dominated region (up to 100 TeV) and with experimental results from IceCube using different methods in the astrophysical dominated region (over PeV). A slight excess is seen in the 100 TeV to 400 TeV energy range, but the results cannot be interpreted before appliction to the full dataset. 
\par
Additionally to unfolding in both angular bins and the overall observed sky, we discussed the approach to treating observables in IceCube to ensure sufficient amount of data available for an analysis, which is not problem-specific and can be expanded to other areas. 
\par
Angular unfolding shows possible separation in up to five zenith bins. The angular fluxes can be discriminated until the last two energy bins, in which the uncertainties surrounding the best estimates overlap. The last two bins correspond to energies of 500 TeV and higher, for which the ratio is expected to return to values of one due to the dominance of the two isotropic components, astrophysical and prompt.  
\par
The angular dependence remains an exciting area to research, and can benefit from introduction of more data with the purpose of separation in the highest energy region, and can be used with aim of advancement in modeling contribution of the three neutrino components to the flux.

\bibliographystyle{ICRC}
\bibliography{references}

%

\clearpage

\section*{Full Author List: IceCube Collaboration}

\scriptsize
\noindent
R. Abbasi$^{17}$,
M. Ackermann$^{63}$,
J. Adams$^{18}$,
S. K. Agarwalla$^{40,\: 64}$,
J. A. Aguilar$^{12}$,
M. Ahlers$^{22}$,
J.M. Alameddine$^{23}$,
N. M. Amin$^{44}$,
K. Andeen$^{42}$,
G. Anton$^{26}$
% ...
\\
\\
% ...
$^{12}$ Universit{\'e} Libre de Bruxelles, Science Faculty CP230, B-1050 Brussels, Belgium \\
% ...
$^{17}$ Department of Physics, Loyola University Chicago, Chicago, IL 60660, USA \\
$^{18}$ Dept. of Physics and Astronomy, University of Canterbury, Private Bag 4800, Christchurch, New Zealand \\
% ...
$^{22}$ Niels Bohr Institute, University of Copenhagen, DK-2100 Copenhagen, Denmark \\
$^{23}$ Dept. of Physics, TU Dortmund University, D-44221 Dortmund, Germany \\
% ...
$^{26}$ Erlangen Centre for Astroparticle Physics, Friedrich-Alexander-Universit{\"a}t Erlangen-N{\"u}rnberg, D-91058 Erlangen, Germany \\
% ...
$^{40}$ Dept. of Physics and Wisconsin IceCube Particle Astrophysics Center, University of Wisconsin{\textendash}Madison, Madison, WI 53706, USA \\
% ...
$^{42}$ Department of Physics, Marquette University, Milwaukee, WI, 53201, USA \\
% ...
$^{44}$ Bartol Research Institute and Dept. of Physics and Astronomy, University of Delaware, Newark, DE 19716, USA \\
% ...
$^{63}$ Deutsches Elektronen-Synchrotron DESY, Platanenallee 6, 15738 Zeuthen, Germany  \\
$^{64}$ Institute of Physics, Sachivalaya Marg, Sainik School Post, Bhubaneswar 751005, India \\
% ...

\subsection*{Acknowledgements}

\noindent
The authors gratefully acknowledge the support from the following agencies and institutions:
USA {\textendash} U.S. National Science Foundation-Office of Polar Programs,
U.S. National Science Foundation-Physics Division,
U.S. National Science Foundation-EPSCoR,
Wisconsin Alumni Research Foundation,
Center for High Throughput Computing (CHTC) at the University of Wisconsin{\textendash}Madison,
Open Science Grid (OSG),
Advanced Cyberinfrastructure Coordination Ecosystem: Services {\&} Support (ACCESS),
Frontera computing project at the Texas Advanced Computing Center,
U.S. Department of Energy-National Energy Research Scientific Computing Center,
Particle astrophysics research computing center at the University of Maryland,
Institute for Cyber-Enabled Research at Michigan State University,
and Astroparticle physics computational facility at Marquette University;
Belgium {\textendash} Funds for Scientific Research (FRS-FNRS and FWO),
FWO Odysseus and Big Science programmes,
and Belgian Federal Science Policy Office (Belspo);
Germany {\textendash} Bundesministerium f{\"u}r Bildung und Forschung (BMBF),
Deutsche Forschungsgemeinschaft (DFG),
Helmholtz Alliance for Astroparticle Physics (HAP),
Initiative and Networking Fund of the Helmholtz Association,
Deutsches Elektronen Synchrotron (DESY),
and High Performance Computing cluster of the RWTH Aachen;
Sweden {\textendash} Swedish Research Council,
Swedish Polar Research Secretariat,
Swedish National Infrastructure for Computing (SNIC),
and Knut and Alice Wallenberg Foundation;
European Union {\textendash} EGI Advanced Computing for research;
Australia {\textendash} Australian Research Council;
Canada {\textendash} Natural Sciences and Engineering Research Council of Canada,
Calcul Qu{\'e}bec, Compute Ontario, Canada Foundation for Innovation, WestGrid, and Compute Canada;
Denmark {\textendash} Villum Fonden, Carlsberg Foundation, and European Commission;
New Zealand {\textendash} Marsden Fund;
Japan {\textendash} Japan Society for Promotion of Science (JSPS)
and Institute for Global Prominent Research (IGPR) of Chiba University;
Korea {\textendash} National Research Foundation of Korea (NRF);
Switzerland {\textendash} Swiss National Science Foundation (SNSF);
United Kingdom {\textendash} Department of Physics, University of Oxford.

\end{document}